%% file: Template_Blind.tex
\documentclass{article}

\usepackage{pgfplots}
\DeclareUnicodeCharacter{2212}{-} 
\usepgfplotslibrary{groupplots,dateplot}
\usetikzlibrary{patterns,shapes.arrows}
\pgfplotsset{compat=newest}

\usepackage{spconf,amsmath,graphicx}
\usepackage{tabularray}
\usepackage[textsize=tiny]{todonotes}
\usepackage[normalem]{ulem} 
\usepackage{subcaption}
\usepackage{cite}
\usepackage{xspace}
\usepackage{algorithmicx,algpseudocode,algorithm}

\algblockdefx{MRepeat}{EndRepeat}{\textbf{repeat}}{}
\algnotext{EndRepeat}

\newcommand{\specaug}{\textsc{SpecAug}\xspace}
\newcommand{\dfilter}{\textsc{DataFilter}\xspace}
\newcommand{\train}{\textsc{Train}\xspace}

\title{CONSISTENCY BASED UNSUPERVISED SELF-TRAINING FOR ASR PERSONALISATION}
\name{BLIND}
\address{BLIND}
\name{\em{Jisi Zhang$^{1*}$, Vandana Rajan$^{1*}$, Haaris Mehmood$^1$, David Tuckey$^1$, Pablo Peso Parada$^1$, Md Asif Jalal$^1$,} \\ \em{Karthikeyan Saravanan$^1$, Gil Ho Lee$^2$, Jungin Lee$^2$, Seokyeong Jung$^2$}}

\address{
    $^1$Samsung Research UK, United Kingdom,\\
    $^2$AI R\&D Group, Samsung Electronics, Suwon, South Korea}
\copyrightnotice{979-8-3503-0689-7/23/\$31.00~\copyright2023 IEEE}
\begin{document}
%
\maketitle
\def\thefootnote{*}\footnotetext{Equal contribution}
\begin{abstract}

On-device Automatic Speech Recognition (ASR) models trained on speech data of a large population might underperform for individuals unseen during training. This is due to a domain shift between user data and the original training data, differed by user's speaking characteristics and environmental acoustic conditions. ASR personalisation is a solution that aims to exploit user data to improve model robustness. The majority of ASR personalisation methods assume labelled user data for supervision. Personalisation without any labelled data is challenging due to limited data size and poor quality of recorded audio samples. This work addresses unsupervised personalisation by developing a novel consistency based training method via pseudo-labelling. Our method achieves a relative Word Error Rate Reduction (WERR) of 17.3\% on unlabelled training data and 8.1\% on held-out data compared to a pre-trained model, and outperforms the current state-of-the art methods.

\end{abstract}
\begin{keywords}
speech recognition, unsupervised, speaker adaptation, personalisation
\end{keywords}
\section{Introduction}
\label{sec:intro}

End-to-end ASR models are known to underperform when deployed in the wild as they face voice characteristics (e.g. accent, tone) and background acoustics (e.g. noise, reverberation) unseen during training~\cite{zhao2018domain, huo2022incremental,sim2019investigation, bell2020adaptation}. A promising approach to remedy this issue consists of fine-tuning a pre-trained ASR model on device with collected user data~\cite{Sim2019AnII,meng2019domain, Huang2021RapidSpeaker}. This type of approach is also known as ``personalisation" or ``adaptation" as it tailors the ASR model to the single user of the device.

While supervised personalisation methods~\cite{Sim2019AnII,meng2019domain, Huang2021RapidSpeaker} substantially improve performance for end-users, they require labelled data which is impractical in many use cases~\cite{Lin2022ListenAB}. This paper focuses on \textit{unsupervised self-training}, which aims to improve the robustness of an ASR model using only unlabelled user data.  Recently, unsupervised ASR personalisation has made progress by incorporating auxiliary speaker features into an ASR model~\cite{Delcroix2018AuxiliaryFB}, exploring self-training methods~\cite{Deng2022ConfidenceSB,Deng2023ConfidenceSB}, and training based on entropy minimisation~\cite{Lin2022ListenAB,kim2023sgem}.

A common pipeline for the unsupervised self-training method contains data filtering, pseudo-labelling, and training~\cite{Deng2022ConfidenceSB,Deng2023ConfidenceSB}. In~\cite{Deng2023ConfidenceSB}, a confidence estimation module uses ASR output probabilities to select a less erroneous subset of utterances from the entire set of unlabelled samples. The filtered samples are processed by the pre-trained model to generate pseudo-labels, which are subsequently used during training. However, the training process can be unstable without accessing labelled data for supervision and the model drifts away due to erroneous pseudo-labels~\cite{Li2020ModelAU}.

Consistency Constraint (CC) forces a model to predict the same results on the same input with various versions of perturbation and has been shown effective for exploring unlabelled data~\cite{SajjadiJT16Regularization,sohn2020fixmatch,Weninger2020SemiSupervisedLW,Sapru2022UsingDA}. Applying various perturbations introduces randomisation to regularise a model, leading to more stable model generalisation~\cite{SajjadiJT16Regularization}. CC has been successfully combined with supervised loss function for semi-supervised learning in computer vision~\cite{sohn2020fixmatch} and speech applications~\cite{Weninger2020SemiSupervisedLW,Sapru2022UsingDA}. However, it has not yet been explored for speech recognition in a fully unsupervised self-training setting.

In this work, we exploit CC to improve the robustness of training process for unsupervised ASR personalisation. We introduce CC to the common unsupervised self-training pipeline, however, perturbations are applied to both pseudo-labelling and the training process, forcing the model to output a consistent label in the vicinity of the training sample after filtering. To the best of our knowledge, our work is the first to apply CC in the context of a fully unsupervised setting for ASR. We compare against a state-of-the-art (SOTA) unsupervised self-training method that combines a data filtering strategy with an adapter based training mechanism~\cite{Deng2022ConfidenceSB} and a separate unsupervised adaptation method based on entropy minimisation~\cite{Lin2022ListenAB}. Our proposed method achieves a 17.3\% relative WERR compared to a pre-trained model, and outperforms the recently developed techniques in literature, leading to a new SOTA performance. The main contributions in this paper are summarised as follows:
\begin{itemize}
    \item We propose a novel consistency based training method for unsupervised personalisation of ASR models which outperforms current SOTA methods.
    \item We empirically show that our proposed method is agnostic to the choice of data-filtering methods and thus can be used in conjunction with any of them. 
    \item We evaluate the robustness of our method on a broad range of English accents with pseudo-label WER ranging from 10\% (high-quality) to 45\% (low-quality).
\end{itemize}
The rest of the paper is organised as follows. Section~\ref{sec:related} reviews related work in the literature. Section~\ref{sec:background} provides background details of an ASR model and a confidence-score based filtering method used in the proposed framework. Section~\ref{sec:proposed} describes the proposed training method based on consistency. Section~\ref{sec:exp_setup} presents implementation details and the experiment setup. Results and analysis are presented in Section~\ref{sec:result}. Finally, the paper is concluded in Section~\ref{sec:conclusion}.

\section{Related work}
\label{sec:related}

Data filtering is a commonly used pre-processing step when exploring unlabelled data~\cite{Khurana2020UnsupervisedDA,Deng2022ConfidenceSB}. Confidence based filtering methods have been developed to select samples with good quality labels that yield lower Word Error Rate (WER) among the whole unlabelled dataset~\cite{kalgaonkar2015estimating, Khurana2020UnsupervisedDA, Gupta2021NeuralUC, dawalatabad2023unsupervised}. For example, in DUST~\cite{Khurana2020UnsupervisedDA, dawalatabad2023unsupervised}, multiple ASR hypotheses are created using dropout, and edit distances from reference (w/o dropout) are used to estimate the confidence of ASR model on the generated transcript. 
Another class of methods use a confidence estimation system that trains a separate neural network using intermediate features derived from the ASR model~\cite{kalgaonkar2015estimating, Gupta2021NeuralUC, Deng2022ConfidenceSB}. For example, both ~\cite{Gupta2021NeuralUC} and~\cite{Deng2022ConfidenceSB} use light-weight binary classifier models to predict whether a given set of ASR features correspond to an error-free pseudo label (WER=0) or not.


Model adaptation methods focus on learning highly compact speaker-dependent parameter representations~\cite{Wang2016AJT, Wang2017UnsupervisedSA, swietojanski2016learning, Deng2022ConfidenceSB}. 
Given a pre-trained model, a diagonal linear transformation of the input features is learned to match the distribution of test data to the training data~\cite{Wang2016AJT}. An ASR model's batch normalization layer parameters can be learned from the target speaker data for adaptation~\cite{Wang2017UnsupervisedSA}.
Additional speaker-dependent parameters can be introduced to an ASR model for adaptation~\cite{swietojanski2016learning,Deng2022ConfidenceSB}.
Specifically, learning hidden unit contributions (LHUC) trains embedding vectors with fixed dimensions to modify the amplitudes of hidden unit activations~\cite{swietojanski2016learning}. Recently, the LHUC method has been successfully applied to a Conformer based end-to-end ASR model for speaker adaptation~\cite{Deng2022ConfidenceSB}. However, the major drawbacks of these techniques are that the selection of adaptation layers in the ASR model is non-trivial and adding new parameters changes the model architecture.

Recently, entropy minimisation based approaches have been shown effective for unsupervised ASR domain adaptation~\cite{Lin2022ListenAB,kim2023sgem}. It aims to reduce uncertainty for samples in a target domain by minimising an entropy loss based on output probabilities from a pre-trained model given the target samples. However, one issue of this method is that the model tends to drift away when the initial predictions are incorrect.

\section{Background}
\label{sec:background}

This section provides the background details of a streaming ASR system and the Neural Confidence Measure (NCM) based data filtering method, which are required for our proposed unsupervised personalisation pipeline.



\subsection{ASR model}
\begin{figure}[t]
  \centering
  \centerline{\includegraphics[width=8.5cm]{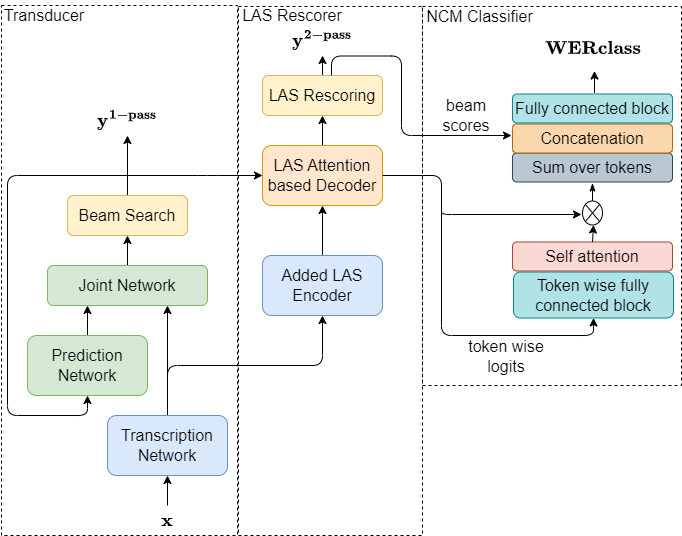}}
\captionsetup{font={small, it}}
\caption{Streaming two-pass end-to-end ASR model architecture. The first pass model is a conformer based transducer. The second pass model is an attention-based encoder-decoder model (LAS). NCM classifier is a confidence estimation module that uses intermediate ASR features for WER based data filtering. }
\label{fig:2pass_e2e_model}
\end{figure}
A state-of-the-art, two-pass Conformer-T model~\cite{Park2023ConformerBasedOS} is used as a pre-trained ASR model for both filtering unlabelled data and adaptation to a target speaker. As shown in Fig.~\ref{fig:2pass_e2e_model}, it 
consists of two sub-models, namely, a parent model and a second-pass model. The parent model is a conformer transducer~\cite{gulati2020conformer}, which consists of a transcription network, a prediction network and a joint network. The transcription network contains a convolution subsampling module followed by stacked convolution-augmented Transformer (Conformer) blocks. The prediction network contains two LSTM layers. The second-pass model is an LAS rescorer, consisting of an LSTM based encoder-decoder architecture. The LAS encoder takes as input the parent's transcription output, and the first-pass prediction is refined by attending to the second-pass encoder outputs.
The model is trained using both a transducer~\cite{Graves2012SequenceTW} loss and a cross-entropy (CE) loss calculated based on the output from the parent model and the second-pass model, respectively.

\subsection{Data filtering methods}
Inspired by recent works that employ WER prediction models for data filtering~\cite{kumar2020utterance,Li2020ConfidenceEF,Gupta2021NeuralUC,Deng2022ConfidenceSB,Deng2023ConfidenceSB}, we use one such existing model that was exclusively developed for our custom ASR model described previously. The NCM binary classification model~\cite{Gupta2021NeuralUC} is used for filtering the transcripts generated by our two-pass ASR model~(see Fig.~\ref{fig:2pass_e2e_model}). The NCM model is made up of dense layers and self-attention mechanism and takes two types of intermediate features from the ASR model as input, namely, second pass decoder output and beam scores. The second pass decoder outputs are logits from the LAS second pass decoder. We use top-K logits corresponding to each decoded token for this decoder. Beam scores are the log-probability scores for each beam assigned by the 2nd pass decoder. Input features are obtained by running the ASR model with beam search and the model is trained using binary cross-entropy loss. The output consists of two classes, WER=0 and WER$>$0 and only the predicted WER=0 samples are selected as the filtered set. Note that different to the original NCM model that used 6 types of features, we use only two based on the relevance of features as shown in the Table 1 of \cite{Gupta2021NeuralUC}. Once the training is complete, the saved NCM model is used to perform pseudo-label filtering on the on-device recorded personal data.

Additionally, we also experiment with two other techniques from literature for data filtering, namely, DUST~\cite{Khurana2020UnsupervisedDA} and Confidence Thresholding (CT)~\cite{Kahn2019SelfTrainingFE}. DUST is based on the intuition that for confident predictions, the ASR output would remain unchanged even if some amount of uncertainty is introduced into the model in the form of dropout. In practice, dropout layers are enabled in the ASR model during evaluation and each utterance is forward propagated through the network for multiple times to generate different hypotheses. Levenshtein edit distances between a reference (hypothesis with no dropout) and each of the hypotheses are calculated. If any of the distances corresponding to an utterance is above a predefined threshold, that utterance is perceived as having a lower confidence by the ASR model and hence rejected. 

Confidence threshold for each utterance is obtained by taking sum of log softmax scores across all tokens. WER values are binarized by taking WER=0 as positive class and WER$>$0 as negative class. The threshold is then found by taking the geometric mean of sensitivity and specificity using an ROC curve.




\section{Proposed method}
\label{sec:proposed}

\begin{figure}[htb]

  \centering
  \centerline{\includegraphics[width=8.5cm]{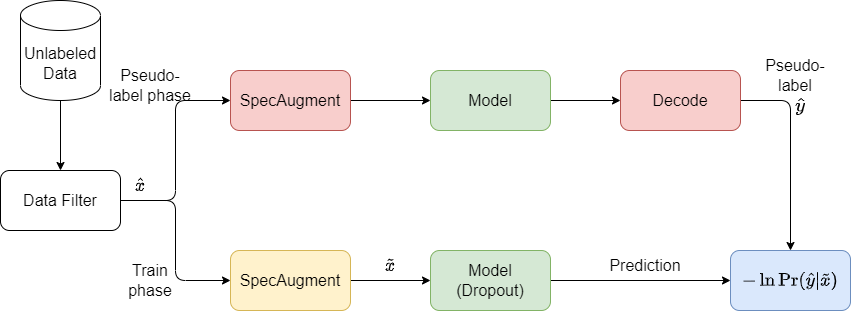}}
\captionsetup{font={small, it}}
\caption{Unsupervised personalisation pipeline based on data filtering and consistency constraint}
\label{fig:unsupervised_perso_pipeline}
\end{figure}

The proposed unsupervised personalisation pipeline starts from filtering unlabelled data to using the filtered data to adapt a pre-trained ASR model based on the Consistency Constraint (CC). The novelty of our adaptation method is that it is the first work to apply CC in the context of fully unsupervised ASR personalisation. The personalisation pipeline~(see Fig.~\ref{fig:unsupervised_perso_pipeline}) based on the proposed CC training is summarised in Algorithm 1. We first apply data filtering $\dfilter$ to the entire unlabelled set $\mathcal{X}$ to obtain filtered set $\mathcal{\hat{X}}$. Subsequently, the model is trained on the filtered set, involving $N$ rounds of pseudo-labelling and training, and in each round the model $f$ is trained for $M$ epochs with the paired audio samples and pseudo-labels $\mathcal{\hat{D}}$.

\begin{algorithm}
\caption{Proposed consistency based unsupervised personalisation pipeline}
\begin{algorithmic}[1]
\State Input: ASR model $f$, weights $\theta$, unlabelled data $\mathcal{X}$
\State $\mathcal{\hat{X}} = \dfilter(\mathcal{X}), \; \theta_{0} = \theta$
\For{$i = 0$, \ldots, $N - 1$}
\hfill// $N$ rounds
\State $\mathcal{\hat{D}} = \{(\hat{x}, f(\specaug(\hat{x}), \theta_{i})),\, \ldots\}\;\; \forall \;\hat{x} \in \mathcal{\hat{X}}$
\State $\theta_{i+1} = \train(f, \theta_{i}, \mathcal{\hat{D}}, M)$
\hfill// $M$ epochs
\EndFor
\State Output: $\theta_N$
\end{algorithmic}
\end{algorithm}



The consistency is realised via pseudo-labelling on utterances augmented with random data perturbations, after which the generated pseudo-labels are used to train the model that are perturbed in a different way. The pseudo-label for each utterance is updated frequently with the latest ASR model during the personalisation process. CC has been commonly applied as a regularisation loss to semi-supervised learning approaches~\cite{sohn2020fixmatch,Weninger2020SemiSupervisedLW,Sapru2022UsingDA,Zhang22MEMO}, in which a main loss is calculated based on the available labelled source domain data. This work uses CC as the only loss for unsupervised personalisation instead of using it as an auxiliary loss. 



Pseudo-labels are generated by decoding the output of the second-pass model via beam search to produce hard labels, a transcription sequence $\hat{y}$. During the training phase, only the parameters of the first-pass model are adapted. Specifically, the first-pass transducer takes augmented input features $\tilde{x}$ and outputs the posterior probabilities of all possible alignments, which are used to calculate the loss against the pseudo-label $\hat{y}$. The loss is then used to train the first-pass transducer comprised of transcription network, prediction network, and joint network. Since the first-pass model is a conformer transducer model, the loss function is implemented as incorporating the CC within the standard RNN-T loss:
\begin{equation}
 	\cal{L}= - \ln \Pr(\hat{\mathit{y}}|\tilde{\mathit{x}})
\end{equation}

SpecAugment~\cite{Park2019SpecAugmentAS,Park2019SpecaugmentOL} is used as the data perturbation for both the pseudo-labelling and the training. Due to randomisation, the SpecAugment applied to a sample during training is different to the pseudo-labelling for the same sample. During training, besides the data perturbation, the model is perturbed by a dropout strategy~\cite{Hinton2012ImprovingNN}. The combination of data as well as model perturbation forms a stronger augmentation for training than that of pseudo-labelling.

\section{Experiment setup}
\label{sec:exp_setup}


\subsection{Data}
The ASR model is pre-trained on 20K hours of English speech data. This includes data from public speech datasets such as LibriSpeech~\cite{panayotov2015librispeech} as well as in-house data from a variety of domains such as search, telephony, far-field, etc. The 
performance of this model on an in-house validation set of 6 hours (5K utterances) is 15.69\% WER.



For personalisation experiments, the proposed method is evaluated on an in-house synthetic user data for a mobile phone use case. There are 12 speakers, each containing three styles of speech: (i) application launch/download commands (Apps), contact call/text commands (Contacts) and common real-world voice assistant commands (Dictations). Table \ref{table:data-perso} provides examples for the three types of data. The average audio length of Apps and Contacts samples is two seconds. The filtered Apps and Contacts data are used to personalise a pre-trained acoustic model. Then, the personalised model is evaluated on the entire (filtered and unfiltered) Apps and Contacts data, and the held-out Dictations data that is unseen during the training. On average, for each speaker, the duration of entire Apps, Contacts and Dictation sets are 8.37, 16.93 and 6.47 minutes respectively. 


\subsection{ASR model configuration}
The transcription network in the two-pass ASR model consists of 16 conformer blocks. Each conformer block consists of one feed-forward layer block, one Convolution Block, one multi-head self-attention block, another one feed-forward layer block, followed by a layer normalisation layer. The prediction network is constructed by stacking two LSTM layers with a dimension of 640. The joint network is a dense layer. In the second-pass model, the encoder contains one LSTM with dimension of 680, and the decoder contains two LSTMs with dimension of 680. When decoding, the beam sizes for the first-pass model and the second-pass LAS re-scoring are set as 4 and 1, respectively. No language model is used for decoding.

Batch size of 16 and Adam optimizer~\cite{Kingma2014AdamAM} are used for all the adaptation experiments. Learning rates for ASR model fine-tuning and LHUC based training are set as 5e-6 and 1e-3, respectively. The SpecAugment used for both pseudo-labelling and training has one frequency mask with size of 13 and two time masks with size of 12. During training, the model is perturbed with dropout of 0.1.

\begin{table}
\centering
\captionsetup{font={small, it}}
\caption{Adaptation data examples: Apps, Contacts, and Dictations}
\begin{tblr}{
  vline{2} = {-}{},
  hline{1,2,4,6,8} = {-}{},
}
Split      & Examples                                        \\
Apps       & Open Messenger                                  \\
          & Install Snapchat                                \\
Contacts   & Send a message to Anne Hathaway                 \\
          & Call Emma Stone                                 \\
Dictations & When does summer start                          \\
          & Set an alarm for seven thirty 
\end{tblr}
\label{table:data-perso}
\end{table}

\subsection{Data filtering method setup}
In the NCM, the first fully connected block consists of two dense layers each with 64 neurons followed by Tanh activation. This is followed by a self-attention layer whose output is summed across all tokens for each utterance and concatenated with the beam scores. This is then passed through another fully connected block made up of two dense layers of 64 neurons each and an output dense layer for the binary class prediction. Training the NCM model uses a batch size of 32 and Adam optimizer~\cite{Kingma2014AdamAM} with an initial learning rate of 1e-3. An exponential decay scheduler with decay rate 0.5 for every 500 training steps was also used. The K value is set as 4. The NCM model is trained using an in-house 6 hours data, where it is split in the ratio 80:20 for training and validation.

The CT method uses the NCM training data (in-house dataset of 6 hours) for finding the threshold, which is used to identify correct pseudo labels from the Apps and Contacts sets of personal data. For DUST, the dropout of 0.2 (which is the same value used during the original ASR training) is enabled for the transformers in ASR model and 5 hypotheses are generated for each utterance. We tried different thresholds for edit distance from \{0.1, 0.3, 0.5, 0.7\} as recommended in~\cite{Khurana2020UnsupervisedDA} and found that the best results are obtained with 0.1 threshold. 

\section{Results and analysis}
\label{sec:result}
This section presents the main results of the proposed method and shows the comparison against recently developed existing unsupervised adaptation methods. The second part of this section conducts a detailed ablation study on the proposed method with respect to data filtering strategies and individual user performance.

\subsection{ASR personalisation results}
Methods used as baselines in this work include noisy student training (NST)~\cite{Park20NST} without accessing labelled source data, confidence score based LHUC~\cite{Deng2022ConfidenceSB}, and a recently proposed unsupervised test-time adaptation approach based on entropy minimization (EM)~\cite{Lin2022ListenAB}. The additional LHUC modules are inserted into the streaming two-pass model used in this work. Different to the strategy used in~\cite{Deng2022ConfidenceSB} that applies only one LHUC layer to the hidden output of the convolution subsampling module, we apply multiple LHUC layers to the transcription network, which yields better results than applying a single layer. Specifically, one LHUC layer is applied to the output of the convolution subsampling module and another to the output of each of the Conformer block. In total, there are 17 LHUC layers. During adaptation, only the LHUC layers are updated and the parameters of the main ASR model are unchanged. The NST method employs data augmentation (SpecAugment) during training, but not for pseudo-labelling. For the EM approach, we calculate and minimise the entropy of the token-wise posterior probabilities output from the second-pass LAS decoder. The NCM data filtering is applied before the EM training. Both NST and EM also update only the first-pass model parameters to make fair comparison with the proposed method.


\begin{table}[b]
\centering
\captionsetup{font={small, it}}
\caption{Word Error Rate (WER) of the proposed method and existing methods for unsupervised personalisation}\vspace*{-5pt}
\begin{tblr}{
  vline{2} = {-}{},
  hline{1-2,7,9} = {-}{},
}
Methods                            & Apps  & Contacts & Dictation \\
Pre-trained                        & 22.66 & 23.49    & 9.43      \\
NST                                & 21.94 & 23.07    & 9.36       \\
EM~\cite{Lin2022ListenAB}          & 20.26 & 23.23    & 9.53      \\
NCM+EM                             & 19.12 & 22.22    & 8.86      \\
NCM+LHUC~\cite{Deng2022ConfidenceSB} & 20.30 & 22.70    & 9.10      \\
NCM+CC+LHUC               & 19.30 & 21.99    & \textbf{8.64}       \\
NCM+CC                    & \textbf{18.73} & \textbf{21.79}    & 8.67      \\
\end{tblr}
\label{table:asr_lhuc}
\end{table}

Table~\ref{table:asr_lhuc} summarises the ASR performance of the baseline and proposed methods. The proposed method achieves 17.3\%, 7.2\%, and 8.1\% relative WERR on Apps, Contacts, and Dictation, respectively, compared to the pre-trained model. It outperforms both entropy minimisation and the confidence score based LHUC, achieving a new SOTA result. The results show that the proposed method not only improves the ASR performance on unlabelled data used for training (Apps \& Contacts), but also generalises well to unseen data (Dictation) spoken by the target speaker. The second half of Table~\ref{table:asr_lhuc} shows the performance of the proposed method (NCM+CC) by adaptation with and without LHUC and the results indicate that addition of LHUC does not provide any improvement.

NCM+CC outperforms the combination of data filtering and entropy minimization (NCM+EM). EM aims to reduce the output uncertainty when a model processes a test sample. When the model prediction is incorrect, the EM approach trains the model to be more confident about its incorrect decision. For the consistency based approach, a model may output different predictions for a highly uncertain sample with various perturbations. Since the model is trained to map this same sample to different predictions, this sample will have lesser impact on the model weight change compared to samples that provide consistent pseudo-labels. 


\subsection{Ablation study}

\begin{table}[b]
\centering
\captionsetup{font={small, it}}
\caption{The effect of three data filtering methods (CT, DUST, NCM) on the unsupervised personalisation performance}\vspace*{-5pt}
\begin{tblr}{
  vline{2} = {-}{},
  hline{1-2,4,7} = {-}{},
}
Methods                            & Apps  & Contacts & Dictation \\
CC                       & 21.25 & 22.71    & 9.04      \\ 
CC (WER=0)                & 20.38 & 22.40    & 8.87     \\ 
CT+CC                     & 18.91 & 22.10  & 8.75 \\
DUST+CC                    & 18.93 & 21.87 & 8.69 \\
NCM+CC                  & \textbf{18.73} & \textbf{21.79} & \textbf{8.67} \\
\end{tblr}
\label{table:asr_filter}
\end{table}
We first investigate the effect of quality of audio samples on the ASR personalisation performance. The Consistency Constraint (CC) based method is tested either on the unfiltered whole data set or the filtered data set based on three filtering strategies, namely Confidence Thresholding (CT), DUST, and the NCM. The method is also tested on training with only audio samples that the ASR system recognises correctly (WER=0), selected based on the ground-truth transcriptions.

\begin{figure*}[!ht]
 \centering
 \resizebox{1.8\columnwidth}{!}{%
 \input{Figures/ablations_rounds_epochs_unsupervised.tex}
 }
 \vspace{-0.15cm}
 \captionsetup{font={small, it}}
 \caption{Word Error Rate Reduction (WERR) compared to the pre-trained model for Apps, Contacts \& Dictation using consistency training (CC) and unsupervised NST for 20 rounds with a choice of 1, 3 or 5 epochs per round. Higher values are better. Plot values are smoothed using an exponential moving average with weight of 0.6. Best viewed in colour.} 
 \label{fig:analysis_werr}
 \vspace{-0.3cm}
\end{figure*}
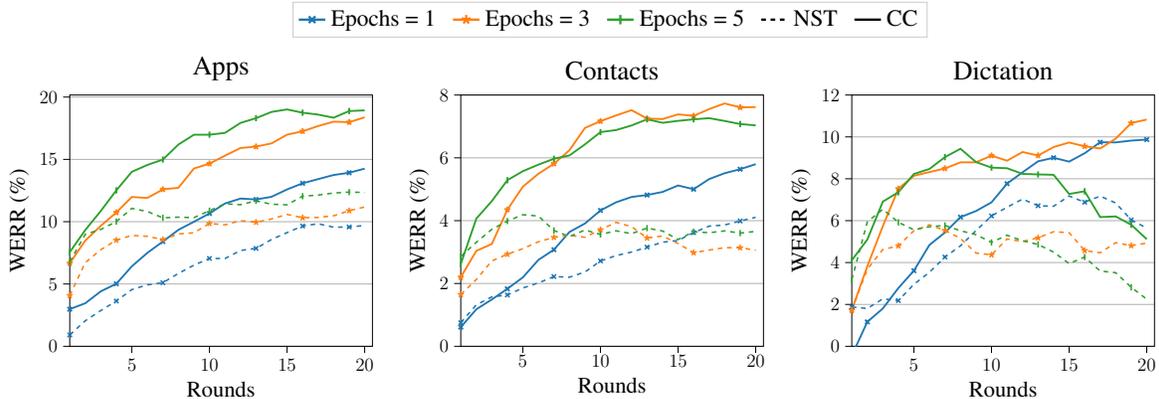

\begin{figure}[!t]
  \centering
  \centerline{\includegraphics[width=9.5cm]{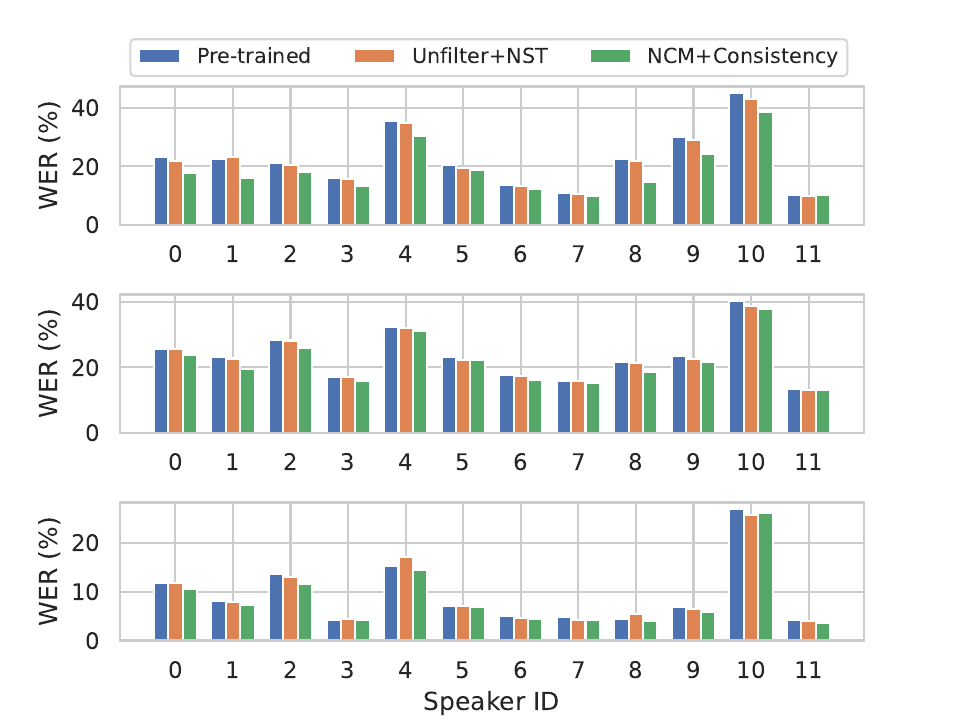}}
\captionsetup{font={small, it}}
\caption{ASR personalisation results for each of the 12 individual users. The pre-trained model, NST trained on unfiltered data, and the proposed method are compared in the plot. (Top: Apps data, Middle: Contacts data, Bottom: Dictation data.)}
\label{fig:analysis_spkr}
%
\end{figure}

In Table~\ref{table:asr_filter}, we first observe that the CC trained with audio samples that the model recognises correctly (WER=0) performs better compared with the CC trained with the whole data (CC). This demonstrates the importance of adapting a given model on samples that yield low WER. The second half of Table~\ref{table:asr_filter} shows the results of CC with different filtering strategies. Remarkably, though trained on filtered data that contains samples with erroneous labels, the CC based training outperforms that trained with WER=0 samples. It suggests that the CC based training is able to explore samples with erroneous labels to adapt the model.

All three data filtering strategies achieve similar performance improvement. Out of the three, NCM is favored for on-device personalisation. NCM is a lightweight model that requires only one-time training and can be easily deployed on device. The CT method requires manual tuning of the threshold value, which is challenging for an end-to-end ASR model because of the well known overconfidence issue that exists in the end-to-end ASR models~\cite{Li2020ConfidenceEF}. However, the NCM can exploit large amounts of training data on server to automatically learn the boundary between reliable and unreliable samples by exploiting multiple intermediate ASR features. Compared to DUST that requires multiple forward passes, which is undesirable due to time and resource constraints, the NCM obtains the result with a single forward pass. 


We perform ablation experiments to study the effect of increasing the number of rounds and epochs per round on the overall WERR. We use epochs $\in \{1, 3, 5\}$ for up to 20 rounds and compare our method against unsupervised NST. Fig.~\ref{fig:analysis_werr} shows that training with five epochs per round can lead to divergence for Dictation which is a classic example of overfitting due to increased model updates. Conversely, training for a single epoch per round leads to sub-optimal convergence due to the increased stochasticity in regenerating pseudo-labels with input augmentation every round. Our method performs up to 40\% better than unsupervised NST which is more susceptible to overfitting due to being easily stuck in a local minima.

We investigate the performance of our proposed method on each individual user, and the analysis is shown in Fig.~\ref{fig:analysis_spkr}. In comparison to the pre-trained model and the baseline NST trained on unfiltered data, our method achieves better or equivalent recognition accuracy for most users on both held-in data (Apps \& Contacts) and held-out data (Dictation). There is a wide range of speech recognition accuracy among the test users, whose WERs range from 10\% to 45\%. The results demonstrate that the proposed method improves the robustness of the training process to erroneous labels.


\section{Conclusions}
\label{sec:conclusion}
This work introduces a novel unsupervised personalisation training method to address a domain shift issue when an ASR model is deployed in the wild. The proposed method performs data filtering of unlabelled user data and applies a consistency constraint to the training process.
A neural confidence measure approach has been employed for data filtering and has demonstrated to effectively discard unreliable audio samples that yield erroneous pseudo-labels. 
To apply consistency constraint, data perturbation is introduced to the iterative pseudo-labelling process, forcing the ASR model to predict the same labels on the same sample with various versions of perturbations. Experiments show that the proposed method reduces the negative impact of low quality samples during training and improves model generalisation to test domain data. We further separately evaluate the consistency based training in combination with three existing filtering methods and all filtering methods achieve similar results. This suggests that the consistency based training can be used in conjunction with a wide range of data filtering strategies.
\bibliographystyle{IEEEbib}
\bibliography{Template_Blind}

\end{document}

%% file: Figures/ablations_rounds_epochs_unsupervised.tex
\begin{tikzpicture}

\definecolor{darkgray176}{RGB}{176,176,176}
\definecolor{darkorange25512714}{RGB}{255,127,14}
\definecolor{forestgreen4416044}{RGB}{44,160,44}
\definecolor{lightgray204}{RGB}{204,204,204}
\definecolor{steelblue31119180}{RGB}{31,119,180}

\begin{groupplot}[group style={
group size=3 by 1, vertical sep=1cm, horizontal sep=2cm
}]
\nextgroupplot[
legend cell align={left},
legend columns=5,
legend style={
  fill opacity=1,
  draw opacity=1,
  text opacity=1,
  at={(1.785,1.3)},
  /tikz/every even column/.append style={column sep=0.25cm},
  nodes={scale=1.5, transform shape},
  anchor=center,
  draw=lightgray204
},
tick align=outside,
tick pos=left,
x grid style={darkgray176},
xlabel={Rounds},
xmin=1, xmax=20.5,
xtick style={color=black},
y grid style={darkgray176},
ylabel={WERR (\%)},
ymajorgrids,
ymin=0, ymax=20.1648029306425,
ytick style={color=black},
every tick label/.append style={font=\large},
label style={font=\Large},
title=Apps,
title style={font=\LARGE}
]

\addlegendimage{mark=x,steelblue31119180,line width=1.5pt, mark options={mark size=3pt}}
\addlegendimage{mark=star,darkorange25512714, line width=1.5pt, mark options={mark size=3pt}}
\addlegendimage{mark=|,forestgreen4416044, line width=1.5pt, mark options={mark size=3pt}}
\addlegendimage{no markers,black,dashed, line width=1.5pt}
\addlegendimage{no markers,black, line width=1.5pt}
\legend{Epochs = 1, Epochs = 3, Epochs = 5, NST, CC}

\addplot [line width=1.2pt, steelblue31119180, mark=x, mark repeat=3]
table {%
1 2.98
2 3.455
3 4.41163265306122
4 5.0175
5 6.41952116585704
6 7.49803168635875
7 8.40830071900761
8 9.34470848608565
9 9.99336202482412
10 10.6600484125401
11 11.4589273568149
12 11.8562212363882
13 11.7816353274595
14 11.9971500828122
15 12.5865671847583
16 13.0800795361134
17 13.4161045994862
18 13.7456962330037
19 13.927428813854
20 14.2564693185546
};
\addplot [line width=1.2pt, darkorange25512714, mark=star, mark repeat=3]
table {%
1 6.66
2 8.485
3 9.66612244897959
4 10.7386764705882
5 11.9840458015267
6 11.9026288936627
7 12.5929005241118
8 12.7219071300617
9 14.2687327160577
10 14.6596030732155
11 15.2980782129395
12 15.9121837017442
13 16.0234555498025
14 16.2862792895082
15 16.9840956764921
16 17.2545336994338
17 17.676791693833
18 18.0261104931019
19 17.9876639530838
20 18.3766125924288
};
\addplot [line width=1.2pt, mark=|,forestgreen4416044, mark repeat=3]
table {%
1 7.56
2 9.32875
3 10.86
4 12.5144117647059
5 14.0045176960444
6 14.5522663802363
7 14.9875448392109
8 16.1927700591568
9 16.9795913815879
10 16.9797558232687
11 17.128392744335
12 17.9267735478314
13 18.3085627719223
14 18.8095302421238
15 19.0098123148976
16 18.7458129118054
17 18.5954622977468
18 18.3452519672978
19 18.8791837160298
20 18.9315121428281
};
\addplot [line width=1pt, steelblue31119180, mark=x, dashed, mark options={solid}, mark repeat=3]
table {%
1 0.91
2 2.0475
3 2.87020408163265
4 3.63297794117647
5 4.5511866759195
6 4.93250268528464
7 5.10431114856857
8 5.84706205215797
9 6.51092746511144
10 7.04981492546908
11 7.04186009539212
12 7.69052806621609
13 7.85853627010397
14 8.57168061166444
15 9.12326771497233
16 9.65411038553512
17 9.83649710329966
18 9.55786996461967
19 9.57472300572828
20 9.69683826816444
};
\addplot [line width=1pt, darkorange25512714, mark=star, dashed, mark options={solid}, mark repeat=3]
table {%
1 4.08
2 6.7175
3 7.74938775510204
4 8.52172794117647
5 8.89832061068702
6 8.82350161117078
7 8.56691827543522
8 9.01568858315281
9 9.10632657198923
10 9.86035526636787
11 9.74379026542899
12 10.0830125723832
13 9.95764380311643
14 10.218790927631
15 10.5834460119238
16 10.3179927198631
17 10.342799830933
18 10.4736931920839
19 10.880240688618
20 11.1801553785271
};
\addplot [line width=1pt, mark=|,forestgreen4416044, dashed, mark options={solid}, mark repeat=3]
table {%
1 6.78
2 8.98625
3 9.38102040816327
4 10.0055514705882
5 11.0701110340042
6 10.814123254565
7 10.3103721456978
8 10.3631090651558
9 10.3254862881845
10 10.8786364604779
11 11.4331937939665
12 11.3637651453239
13 11.67065991306
14 11.4021855601535
15 11.3532883453559
16 12.0441679119326
17 12.1345160400795
18 12.3007265043753
19 12.3724402725735
20 12.3474632503446
};

\nextgroupplot[
tick align=outside,
tick pos=left,
x grid style={darkgray176},
xlabel={Rounds},
xmin=1, xmax=20.5,
xtick style={color=black},
y grid style={darkgray176},
ylabel={WERR (\%)},
ymajorgrids,
ymin=0, ymax=8,
ytick style={color=black},
xticklabel style = {font=\large,yshift=0.5ex},
label style={font=\Large},
every tick label/.append style={font=\large},
title=Contacts,
title style={font=\LARGE},
]
\addplot [line width=1.2pt, steelblue31119180, mark=x, mark repeat=3]
table {%
1 0.61
2 1.17875
3 1.49061224489796
4 1.83040441176471
5 2.19455933379597
6 2.7506820622986
7 3.07550212015065
8 3.63876187302116
9 3.90999048329353
10 4.32450067459421
11 4.58364055661944
12 4.75455638051807
13 4.81681514257494
14 4.91416537323947
15 5.12059628465858
16 5.0003238406168
17 5.32424913410137
18 5.5105684030377
19 5.63834882824729
20 5.79901517116443
};
\addplot [line width=1.2pt, darkorange25512714, mark=star, mark repeat=3]
table {%
1 2.2
2 3.0375
3 3.25816326530612
4 4.34816176470588
5 5.08195697432339
6 5.49232008592911
7 5.81235086517949
8 6.24263768538577
9 6.95273879433673
10 7.16895062687399
11 7.35002731702536
12 7.51838286359516
13 7.25068007974238
14 7.23439528640326
15 7.38470784655073
16 7.33481063137763
17 7.55292329801487
18 7.73177214268983
19 7.60705568589513
20 7.61623374710154
};
\addplot [line width=1.2pt, mark=|,forestgreen4416044, mark repeat=3]
table {%
1 2.62
2 4.06375
3 4.6230612244898
4 5.28801470588235
5 5.57513532269257
6 5.78276852846402
7 5.97092838894888
8 6.07632733086152
9 6.42935409492501
10 6.81997438849704
11 6.88421770596266
12 7.03084980975102
13 7.22676576568594
14 7.11597263730263
15 7.17761256454982
16 7.22658135336342
17 7.26395513815133
18 7.17436398401706
19 7.07661243382432
20 7.03396590106978
};
\addplot [line width=1pt, steelblue31119180, mark=x, dashed, mark options={solid}, mark repeat=3]
table {%
1 0.75
2 1.31875
3 1.57448979591837
4 1.63216911764706
5 1.85676613462873
6 2.01756176154672
7 2.22020990808291
8 2.19978284869189
9 2.36554016101854
10 2.71745197276568
11 2.88708661314065
12 3.00049884162003
13 3.15650305730464
14 3.31402527535222
15 3.3684407505507
16 3.62113573836379
17 3.82471590233933
18 3.86283341261629
19 3.98970777878285
20 4.10582891284235
};
\addplot [line width=1pt, darkorange25512714, mark=star, dashed, mark options={solid}, mark repeat=3]
table {%
1 1.65
2 2.125
3 2.71938775510204
4 2.93106617647059
5 3.10843164469119
6 3.32726906552095
7 3.46830967368116
8 3.53409067238794
9 3.47182692834851
10 3.71255172781834
11 3.94437207408246
12 3.81033146722416
13 3.44972790705845
14 3.50988388496431
15 3.27381933697273
16 2.9802087714161
17 3.06413946951309
18 3.13849123286972
19 3.13909477649935
20 3.05945395410496
};
\addplot [line width=1pt, mark=|,forestgreen4416044, dashed, mark options={solid}, mark repeat=3]
table {%
1 2.86
2 3.31
3 3.72326530612245
4 3.98830882352941
5 4.18855655794587
6 4.14720461868958
7 3.68745265874793
8 3.49320904849192
9 3.70202986176984
10 3.56036130145172
11 3.67663863160161
12 3.58578541127579
13 3.76370362029426
14 3.67014885876972
15 3.397961336776
16 3.64684701545099
17 3.6161030053367
18 3.67366764946479
19 3.60419635637796
20 3.65851979997795
};

\nextgroupplot[
tick align=outside,
tick pos=left,
x grid style={darkgray176},
xlabel={Rounds},
xmin=1, xmax=20.5,
xtick style={color=black},
y grid style={darkgray176},
ylabel={WERR (\%)},
ymajorgrids,
ymin=0, ymax=12,
ytick style={color=black},
every tick label/.append style={font=\large},
label style={font=\Large},
title=Dictation,
title style={font=\LARGE},
]
\addplot [line width=1.2pt, steelblue31119180, mark=x, mark repeat=3]
table {%
1 -0.44
2 1.16625
3 1.80591836734694
4 2.78205882352941
5 3.60958362248439
6 4.83911519871106
7 5.43206800284443
8 6.16759482794534
9 6.45949861438823
10 6.87420675014547
11 7.76776585621773
12 8.30983949004156
13 8.8345890533731
14 9.0048868844957
15 8.81884465643529
16 9.21941980051194
17 9.74374063030842
18 9.73824381992923
19 9.82295145369971
20 9.87377273032622
};
\addplot [line width=1.2pt, darkorange25512714, mark=star, mark repeat=3]
table {%
1 1.74
2 3.82125
3 5.73897959183673
4 7.52253676470588
5 8.14166551006245
6 8.32138426423201
7 8.49365298006268
8 8.78508675637394
9 8.78707206112208
10 9.10617271631109
11 8.86282075603907
12 9.27859750911526
13 9.11093953251598
14 9.51488026428749
15 9.73303072964568
16 9.55176730138208
17 9.44704265444491
18 9.91627324770566
19 10.6618093785937
20 10.8290917432645
};
\addplot [line width=1.2pt, mark=|,forestgreen4416044, mark repeat=3]
table {%
1 4.11
2 5.0975
3 6.91
4 7.35577205882353
5 8.22939625260236
6 8.46880773361976
7 9.03308067107377
8 9.43459225545742
9 8.79834343104164
10 8.53340407163144
11 8.50393553198458
12 8.23377323467411
13 8.20823058032519
14 8.18892321814213
15 7.27292324145037
16 7.40379086406489
17 6.17006568998023
18 6.20204266156799
19 5.80520141503291
20 5.11509561765839
};
\addplot [line width=1pt, steelblue31119180, mark=x, dashed, mark options={solid}, mark repeat=3]
table {%
1 1.88
2 1.805
3 2.26673469387755
4 2.18551470588235
5 2.959493407356
6 3.53452470461869
7 4.26105296426032
8 4.79764445509082
9 5.52188535265087
10 6.22538500496983
11 6.64073789102451
12 7.02928852477089
13 6.71316022936893
14 6.68787632410034
15 7.17695575226359
16 6.88209026656385
17 7.17330345260257
18 6.84394862229925
19 6.03431983786099
20 5.64457765312481
};
\addplot [line width=1pt, darkorange25512714, mark=star, dashed, mark options={solid}, mark repeat=3]
table {%
1 1.72
2 3.6825
3 4.62
4 4.79922794117647
5 5.51521165857044
6 5.80253222341568
7 5.53400221233111
8 5.1499507373771
9 4.45496660360125
10 4.36845687088835
11 5.12381453791124
12 5.00603233653129
13 5.17984641516845
14 5.47213690009631
15 5.4232591584776
16 4.57371582957636
17 4.46421096222985
18 4.94657556626619
19 4.80793689167704
20 4.92876655273008
};
\addplot [line width=1pt, mark=|,forestgreen4416044, dashed, mark options={solid}, mark repeat=3]
table {%
1 3.18
2 5.9425
3 6.54326530612245
4 5.92136029411765
5 5.56511450381679
6 5.70982008592911
7 5.77162211277621
8 5.51872557698717
9 5.33336736245514
10 4.96177353167585
11 5.30631410205156
12 5.03920702726873
13 4.86329446234283
14 4.49769017466176
15 3.95035675684244
16 4.26230205730323
17 3.59326798926367
18 3.5159529414453
19 2.81752920573658
20 2.25049679183849
};
\end{groupplot}

\end{tikzpicture}